# Computing the Fewest-turn Map Directions based on the Connectivity of Natural Roads

Bin Jiang and Xintao Liu

Department of Technology and Built Environment, Division of Geomatics
University of Gävle, SE-801 76 Gävle, Sweden
Email: bin.jiang@hig.se, xintao.liu@hig.se

**Abstract**
In this paper, we introduced a novel approach to computing the fewest-turn map directions or routes based on the concept of natural roads. Natural roads are joined road segments that perceptually constitute good continuity. This approach relies on the connectivity of natural roads rather than that of road segments for computing routes or map directions. Because of this, the derived routes posses the fewest turns. However, what we intend to achieve are the routes that not only possess the fewest turns, but are also as short as possible. This kind of map direction is more effective and favorable by people, because they bear less cognitive burden. Furthermore, the computation of the routes is more efficient, since it is based on the graph encoding the connectivity of roads, which is substantially smaller than the graph of road segments. We made experiments applied to eight urban street networks from North America and Europe in order to illustrate the above stated advantages. The experimental results indicate that the fewest-turn routes posses fewer turns and shorter distances than the simplest paths and the routes provided by Google Maps. For example, the fewest-turn-and-shortest routes are on average 15% shorter than the routes suggested by Google Maps, while the number of turns is just half as much. This approach is a key technology behind FromToMap.org - a web mapping service using openstreetmap data.

## 1. Introduction
Map directions refer to the turn-by-turn route instructions advising a traveler from a point of origin to a final destination (sometimes passing multiple locations in between) in a road network. Nowadays, many web mapping services such as Google Maps and Bing Maps provide this kind of map directions for route planning. The map directions can be done with respect to different modes of transport: driving, walking, cycling and public transport. Conventionally, the map directions are computed based on the least cost of distance, i.e. the shortest distance routes. The classic Dijkstra algorithm and A* algorithm are commonly used for this computation (Zhan and Noon 1998), but recent efforts have been applied toward dealing with the computation efficiently for large scale networks (e.g., Sanders and Schultes 2005). In fact, human travelers have many different criteria for selecting routes. Gollege and Gärling (2004) summarized over 20 different types of route selection criteria, among which the criterion of fewest turns is probably the most often used one. The routes with fewest turns imply fewer slow-down-and-speed-ups, saving time and petrol alike, as well as reducing vehicle emission making it also environmentally responsible. The routes with fewest turns are also called simplest paths (Duckham and Kulik 2003, Mark 1985), as they bear less cognitive burden. This type of path is particularly favored when navigating in an unfamiliar environment.

To compute map directions we need a road network represented as a graph $G(V, E)$, in which nodes $V$ represent road junctions and links $E$ are road segments linking two adjacent junctions, i.e., $G^G(J, S)$, where $J$ is the set of junctions, $S$ is the set of road segments, and superscript $G$ indicates a geometry oriented representation. Instead of road segments, the links can also be part of a road segment called arcs. This road representation fairly well reflects the connectivity of individual road segments via common junctions. Even though this road representation forms a graph – topology in essence, it is still geometry oriented in the sense that (1) every junction or node has a precise geometric location, and (2) every road segment has a weight based on the geometric distance (Jiang and Liu 2009). This road representation is commonly used to compute map directions: either for the shortest distance paths or the simplest paths. We will now refer to the shortest distance as the shortest geometric distance to differentiate it from the shortest topological distance which will be introduced. Furthermore, the reason it is called geometry oriented representation lies in the fact that this representation is hard, if not impossible, to incorporate topological or semantic information (*c.f.* section 2 for a more detailed account) in the course of computing routes or map directions. Therefore, map directions based on this representation are detailed turn-by-turn directions at the level of road segments, and it is inevitable for this representation to avoid unnecessary turns.

The computation of map directions based on the connectivity of road segments violates a basic principle of how human beings conceptualize and choose routes in daily life. This basic principle is that people communicate routes based on the connectivity of roads. We often hear map directions as such: from where you are currently at road A, head west at the third intersection, turn left to the $5^{th}$ avenue, after 4 blocks turn right to get to road B. Fundamental to this orally provided direction is a chain of connectivity of roads rather than that of road segments. For example, road A connects to one road, which connects to another road… which connects to road B. The number of intermediate roads plus one indicates the fewest turns (Jiang 2004, Turner and Dalton 2005) or the shortest topological distance. An empirical study using massive GPS data (Turner 2009) has proved that most of the time people indeed follow the fewest-turn routes rather than the shortest distance paths.

In this paper, we introduce an alternative representation that is based on the connectivity of individual natural roads. This connectivity is represented by a unit graph, $G^T(R, J)$, whose links value is always set to 1, and where $R$ is the set of roads that are represented by individual nodes, $J$ is the set of junctions that are represented by links, and superscript $T$ indicates a topology oriented representation. By roads, we refer to either natural roads or named roads. Natural roads are joined road segments that perceptually constitute good continuity. The join process is self-organized in terms of the smallest deflection angle among adjacent segments, so they are also referred to as self-organized natural roads (Jiang, Zhao and Yin 2008). Named roads are identified by unique names (Jiang and Claramunt 2004). Both natural and named roads have shown many modeling advantages in structuring road networks and in predicting traffic flow (Thomson 2003, Jiang and Claramunt 2004, Turner 2007, Tomko, Winter and Claramunt 2008, Jiang and Liu 2009). In this paper, we employ the concept of natural roads for another promising application - computing map directions. What we aim to achieve is identifying the routes that not only possess the fewest turns, but are also as short as possible. Experiments carried out indicate that the fewest-turns routes are superior to the simplest paths and the Google Maps routes in terms of the number of routes and distances involved. This approach is a key technology behind the FromToMap service we have been developing for routing planning and for personal navigation (*c.f.*, the site [www.fromtomap.org](www.fromtomap.org)).

The remainder of this paper is organized as follows. Section 2 presents an example to provide motivation for the basic idea of map directions based on the connectivity of roads. Section 3 describes in details a set of algorithms for computing routes with the fewest turns and the shortest distances. To demonstrate the advantages of our approach or algorithms, section 4 discusses our experimental results in comparison with the simplest paths and Google Maps' directions. Finally section 5 draws a conclusion.

## 2. Map directions based on a notational street network
To motivate the approach, let us start with a notational street network illustrated in Figure 1(a). It is a Manhattan like street network, which involves 8 streets intersected at 16 junctions. Ignoring the 16 street segments without a node in one end, it is a graph about the connectivity of 24 street segments via 16 nodes, or inversely, the connectivity of 16 nodes via 24 street segments. It is rather obvious that the shortest geometric distance between location (F) and location (T) is 6 blocks. There are many routes with the distance of 6, and the blue (dashed line) is one of the many. Out of the many shortest geometric distance routes, there are only two routes that have the fewest turns, and the red (dotted line) is one of the two. The shortest route with the most number of turns is the one along the diagonal direction between F and T. It involves in total 5 turns, forming a zigzag like path.

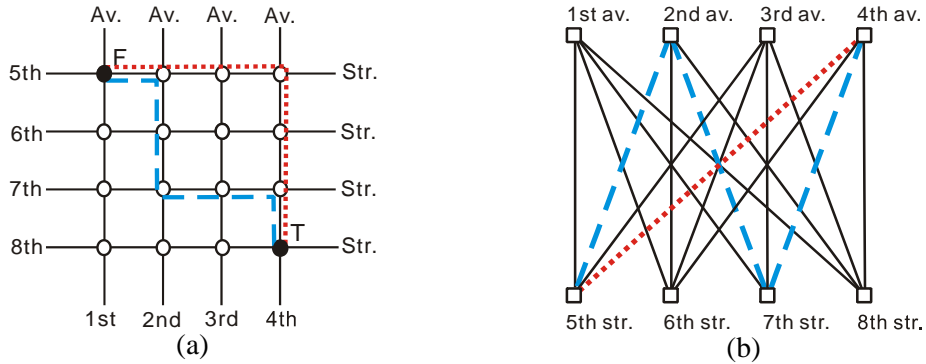

Figure 1: (Color online) Routes with shortest distance (dashed blue lines) and fewest turns (dotted red lines) shown in (a) geometry oriented representation, (b) topology oriented representation

The very reason there are so many turns involved in the routes lies in the geometry oriented representation that lacks related topological or semantic information. The two types of information are higher order information, which is essential in personal navigation. For example, it is not difficult to note that the $5^{th}$ street and the $4^{th}$ avenue are directly intersected, so from F to T needs only one turn, or alternatively, one turn from the $1^{st}$ avenue to the $8^{th}$ street. The topological information can be intuitively reflected in a topology oriented representation as shown Figure 1b. The topological representation, based on the connectivity of streets, is also called connectivity graph, in which each street is collapsed into one node and the corresponding street-street intersections are represented as links. Unlike the geometry oriented graph, the connectivity graph is a unit graph, whose links value is always set to 1. We can note that the routes with the fewest turns are those with the shortest topological distances in the topological representation. There are semantic meanings attached to the streets as well. For example, streets are narrower than avenues, and streets are horizontal (east-west), while avenues are vertical (south-north). Of course these meanings are not absolute, and they may vary from one place to another.

Let us relocate the locations of F and T to the $5^{th}$ street and the $8^{th}$ street respectively as shown in Figure 2a. Now it is clear that the $5^{th}$ and $8^{th}$ streets are not directly intersected, instead they are parallel. However, they are connected via one of the four avenues – the four vertical streets. This implies that the topological distance between $5^{th}$ street and $8^{th}$ street is 2 as shown in Figure 2b. Eventually, the least number of turns between locations F and T is two. Note that there are four routes between F and T that are with the fewest turns, and two of which are indicated as red (dotted lines). To this point, we have seen how the number of turns can be computed from the connectivity graph or based on the connectivity of streets.

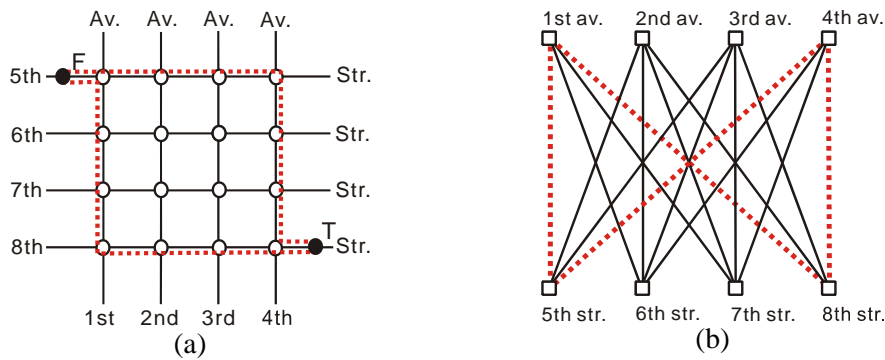

Figure 2: (Color online) Two of the four routes (red dotted lines) with the least turns between locations F and T shown in (a) geometric representation, and (b) topological representation

In our approach, not every junction is a decision point. This is fundamentally different from the conventional approach using the geometric representation for modeling turn cost (Winter 2002). We can achieve much simpler map direction, with only 50% turns on average of the corresponding shortest geometric distance path. It would help greatly for the spatial chunk process as suggested by Klippel, Tappe and Habel (2003), since the number of decision points is greatly reduced when compared to conventional approaches. Due to the simplicity of the example used, there is little ambiguity in concepts such as turns, natural roads and named roads. The ambiguity in the concepts with real world road networks will be discussed and dealt with in the following algorithms to achieve fewest-turn routes as well as fewest-turn-and-shortest routes. We will further see through the following experiments urban morphological structure has an impact on the derivation of map directions.

### 3. Algorithms for computing map directions
We have seen from the above example that we can indeed achieve routes with the fewest turns based on the connectivity of natural roads. Interestingly, the routes with the fewest turns have the same distance as the shortest geometric distance paths. However, this is not true, because real world road networks are unlikely to be laid out in a grid. Before getting into detail on the algorithms, let us clarify a few related concepts. First, turns refer to the change from one natural road to another, not the direction changes within a natural road. For example, while driving in a ring road, we turn our wheels continuously, but these are not deemed turns. Second, we adopted the every-best-fit join principle for generating natural roads, since the generated natural roads match fairly well to named roads (Jiang, Zhao and Yin 2008). The natural roads that match well with named roads are more "natural". According to the same study, we chose a deflection angle of 45 degrees as a threshold for the termination of the join process. The connectivity graph based on natural roads possesses turn information (or topological information) in terms of what roads are connected to what other roads. Third, named roads, unlike natural roads, bear a certain degree of arbitrariness (because of the tradition of how roads are named) and incompleteness (due to missing names for some segments). This is the major reason we did not rely on the connectivity of named roads for computing map directions. However, named roads possess much semantic information such as road hierarchy, names, and other meaningful elements that are essential for human navigation. Importantly, people, most of time, rely on named roads, in particular those predominant ones, for communicating routes or places (Tomko, Winter and Claramunt 2008).

In what follows of this section, we will present details on our algorithms. We take a two-step approach: first a set of procedures for computing the route with the fewest turns (not necessarily the shortest in geometric distance), and second an additional function for splitting natural roads into straighter and shorter ones in order to re-compute the route that is not only with the fewest turns but also as short as possible, i.e., fewest-turn-and-shortest route. Furthermore, we present a function to automatically generate turn-by-turn instructions.

### 3.1 Computing the fewest-turn routes
We developed an algorithm that consists of three sequentially run functions for computing the fewest-turn route from extracted natural roads. First of all, the connectivity of natural roads in terms of what roads are connected to what other roads, or a connectivity graph, is used to compute the shortest topological distance ($D_t$) between the start and end road. We use the Breadth-First-Search strategy to traverse the connectivity graph from the start road till the end road to get the shortest topological distances between the start road and all other roads including the end road. This is the major task of the first function. The resulting shortest topological distance ($D_t$) is then used as a variable for the second function to obtain all possible shortest topological paths. The process goes like this: Use the Depth-First-Search strategy to traverse the connectivity graph from the start road, and continuously compare whether or not the current topological distance equals the shortest topological distance ($D_t$). If true, and if the current node is the end node, then a fewest-turn path is formed. Continue the process till all fewest-turn paths are exhausted. Usually there exist multiple shortest topological paths. Finally, these shortest topological paths are further processed by the third function in order to select the only one with the shortest geometric distance. This path is supposed to be the fewest-turn route.

**Algorithm I: Computing the fewest-turn routes**
--------------------------------------------------------------------------------
// Recursive function for calculating shortest topological distance

Input: Connectivity graph of natural roads
Output: Shortest topological distance between start and end road

```
topDistance = 0;
Set all adjacent nodes of the start node as "current unvisited adjacent nodes";
Function ShortestTopDistance (topDistance, current unvisited adjacent nodes)
    topDistance += 1;
    For each node in unvisited adjacent nodes
        Current node = topDistance;
        Search unvisited adjacent nodes of current node and add them to the next unvisited nodes;
    If (next unvisited nodes include end node) Then
        return;
    ShortestTopDistance(topDistance, next unvisited adjacent nodes);
```

// Recursive function for obtaining all possible shortest topological (or fewest-turn) routes
--------------------------------------------------------------------------------
Input: Connectivity graph of natural roads
Variables: Shortest topological distances of all nodes between start end road
Output: All possible shortest topological routes between start and end road

```
topDistance = 0;
currentNode = start node;
Add currentRoad to currentRoute;

Function SearchAllPossibleRoutes (currentNode, topDistance, currentRoute)
    topDistance += 1;
    Select nodes adjacent with currentNode into "current adjacent nodes";
    For each node in current adjacent nodes
        If (adjacentNodeTopDistance == topDistance) Then
            Add adjacentNode to currentRoute;
            If (adjacentNode == endNode) Then
                Add currentRoute to all possible shortest topological routes;
            Else
                SearchAllPossibleRoutes(adjacentNode, topDistance, currentRoute);
```

// Loop procedure for selecting the fewest-turn route with the shortest distance
--------------------------------------------------------------------------------
Input: Connectivity graph of natural roads
Variables: All possible shortest topological routes between start to end road.
Output: fewest-turn route

```
Function FewestTurnRoute ()
    fewest-turn route = null;
    shortestDistance = positive infinity;
    For each route in shortest topological routes
        If (currentRouteDistance < shortestDistance) Then
            shortestDistance = currentRouteDistance;
            fewest-turn route = currentRoute;
```

### 3.2 Computing fewest-turn-and-shortest routes

The above functions or algorithm can guarantee to obtain the fewest-turn routes between two locations F and T. However, the fewest-turn routes are usually longer than the corresponding shortest geometric distance path. In fact, that the fewest-turn routes are the same distance as the shortest geometric distance path is occurring only when the road networks are perfect grids like the above example. None of the real world road networks are perfect grids, so the fewest-turn routes are always equal to or longer than the corresponding shortest geometric distance paths. This is inevitable. The more a road network deviates from a grid, the longer the fewest-turn route is. We recognized this issue and suggested some solutions in order to derive fewest-turn-and-shortest routes.

A major reason the fewest-turn routes are longer is due to the fact that natural roads can sometimes be greatly curved and very long, thus creating big bends. Any two locations along a big bend are clearly with the fewest turn (i.e., 0 turn), but the two locations are very likely to link each other with some short cut routes with more turns. This is illustrated in Figure 3a, where the red route along the bend is with 0 turn, while the blue route with 1 turn. In order to compute fewest-turn-and-shortest routes, we must split natural roads into straighter and shorter ones. This is done through the Douglas-Peucker algorithm (Douglas and Peucker 1973). This algorithm was initially developed for line simplification by identifying critical points, where big bending is occurring or big bends are formed. A line gets simplified by keeping critical points and removing the rest. Here we adopt the algorithm to split a natural road at critical points into multiple straighter and shorter ones; the details of the function are shown below. The split natural roads are then sent to the three above functions to generate the fewest-turn-and-shortest routes. A similar case is the situation in which two roads intersect at a sharp angle, and there is a short cut between F and T indicated by the blue line (Figure 3b).

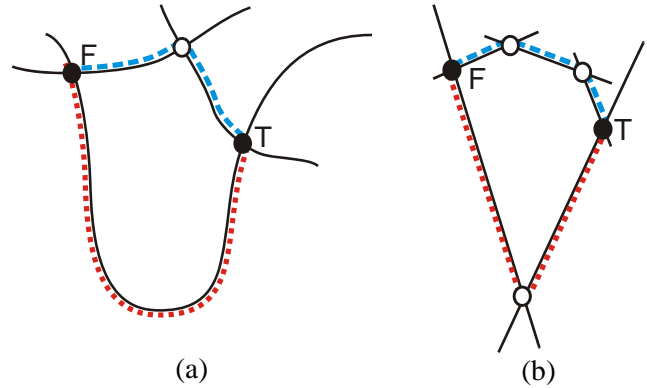

(a)          (b)

Figure 3: (Color online) Routes with fewest turns (red dotted lines) are longer than the shortest path (blue dashed lines) through (a) a bend of one road, and (b) a sharp-angled intersection of two roads

**Algorithm II: Computing fewest-turn-and-shortest route**
--------------------------------------------------------------------------------
```
Input: Point lists of natural roads
Variables: Distance and Ratio
Output: Splitted natural roads

Recursive function SplitNaturalRoad (PointList, Distance, Ratio, )
    SegmentLine = segment starts from the first point and end at the last point;
    MaxOrthogonalDistance = negative infinity;
    RoadLength = Length (PointList);
    SplitIndex = -1;
    For i = 2 to PointCount
        d = OrthogonalDistance (PointList[i], SegmentLine);
        If (pMaxOrthoDist < d)
            Point Index = i;
            MaxOrthogonalDistance = d;

    If (MaxOrthogonalDistance >= Distance or MaxOrthogonalDistance/ StraightlineDistance >= Ratio)
        SplitNaturalRoad (PointList [1...SplitIndex], Distance, Ratio);
        SplitNaturalRoad (PointList [SplitIndex...PointCount], Distance, Ratio);
        Add PointList [1...SplitIndex] as splitted natural road;
        Add PointList [SplitIndex...PointCount] as splitted natural road;

Call ShortestTopDistance ();

Call SearchAllPossibleRoutes ();

Call FewestTurnRoute ();
```

### 3.3 Generating turn-by-turn instructions

The above functions can generate the fewest-turn-and-shortest routes. Next, we need to consider how to generate turn-by-turn map directions in some well structured formats that can be easily remembered by human navigators. In this respect, we adopt a hierarchy of natural roads, named roads, and road segments to formulate the map directions (Algorithm III). In other words, we first consider turn information or topological information, then semantic information, and finally geometric information. This kind of structure conforms well to human spatial cognition, because topological information, and semantic information alike, is higher order information in human navigation.

**Algorithm III: Generating turn-by-turn instructions**
------------------------------------------------------------------------------
```
Input: Fewest-turn routes or fewest-turn-and-shortest routes
Output: XML route instruction

Function RouteInstructions (Route)
    Build XML document;
    For each natural road in Route
        Add parts of the current natural road belonging to the route to XML document;
        For each named road in current natural road
            Add current named road to XML document;
            For each segment in current named road
                Add segment to XML document;
```

Let us add a few words about computational complexity and efficiency of the algorithms. As we have shown, we adopted the Dijkstra algorithm to traverse from one single source node to all other nodes until the destination is reached. The traversal process is done through the Breadth-First-Search strategy. In this regard, the computational complexity applied to $G^G$ and $G^T$ would be the same. However, our approach to computing map directions is more efficient, because $G^T$ is just half or one-third of $G^G$ according to the experiments that follow (*c.f.*, Table 1).

## 4. Results and discussion

We carried out some experiments applied to eight urban street networks from North America and Europe, in order to illustrate how our approach is superior to existing solutions. The eight cities were deliberately and carefully chosen because of their different morphological structures. Six of them were taken from typical street patterns (Jacobs 1995) and two from previous studies, US cities being grid-like and planned, while European cities being irregular and self-evolved. All road networks are taken from openstreetmap (OSM) with the exception of Bloomington. For Bloomington and for a comparison purpose, we adopted the same data set used in Duckham and Kulik (2003), originally taken from the US Census Bureau TIGER data. The sizes and maps of the eight networks are respectively shown in Table 1 and Figure 4. Note that Arcs and ArcsX indicate the size of geometry oriented representation, while Roads and RoadsX indicate the size of topology oriented representation. Roads(I) are for computing fewest-turn paths and roads(II) for fewest-turn-and-shortest paths. We can also note from Table 1 that the size of topological representation is just one third or half of the corresponding geometric representation. The experiments were done through comparisons between our solutions: fewest-turn and fewest-turn-and-shortest routes and existing solutions: shortest paths, simplest paths, and Google Maps routes (Table 2).

Table 1: Size of the eight street networks from both geometric and topological perspectives
(Note: ArcsX = Arc-arc intersections, RoadsX = Road-road intersections, Roads(I) = Natural roads before split, Roads(II) = Natural roads after split)

|  | Arcs | ArcsX | Roads(I) | Roads(I)X | Roads(II) | Roads(II)X |
|---|---|---|---|---|---|---|
| Bloomington | 4696 | 3252 | 1477 | 2226 | 1960 | 2834 |
| Manhattan (NY) | 8310 | 4235 | 806 | 4271 | 956 | 4491 |
| San Francisco | 17991 | 10689 | 3031 | 9921 | 3814 | 10997 |
| Toronto | 13084 | 7126 | 2601 | 6192 | 3213 | 6880 |
| Gävle | 4258 | 2602 | 947 | 1667 | 1532 | 2332 |
| Copenhagen | 6699 | 4236 | 1650 | 3575 | 2041 | 4042 |
| London | 8511 | 5265 | 2206 | 4184 | 2690 | 4738 |
| Paris | 20161 | 11507 | 4543 | 11794 | 5345 | 12986 |

Before reporting the experimental results, it is important to point out that we disagree with Duckham and Kulik (2003) on their approach to comparing two variables. This disagreement is documented in the Appendix at the end of this paper. Even with their approach, we were unable to replicate their result – the simplest paths are on average 16% longer than the corresponding shortest paths. Instead, we ended up with a result of 23.5% (*c.f.,* the first row of Table 3). Let us take a look at how their comparison leads to some contradictory result. We note that the simplest paths are 23.5% longer than the shortest paths (*c.f.*, column SP/ST), and the fewest-turn paths are 21.4% longer than the shortest paths (*c.f.*, column FT/ST). This implies that the simplest paths are longer than the shortest paths, but column FT/SP indicates that the simplest paths are shorter than the shortest paths. While communicating with the first author on the inconsistent results, Duckham advised that they pre-processed that data before the computation, and corrected some road connectivity which he could not figure out how they did in detail. Nevertheless, we provide the result using their comparison approach in Table 3 – the row marked with *. The rest experimental results are based on our comparison approach based on Equation A2 (*c.f.*, the Appendix).

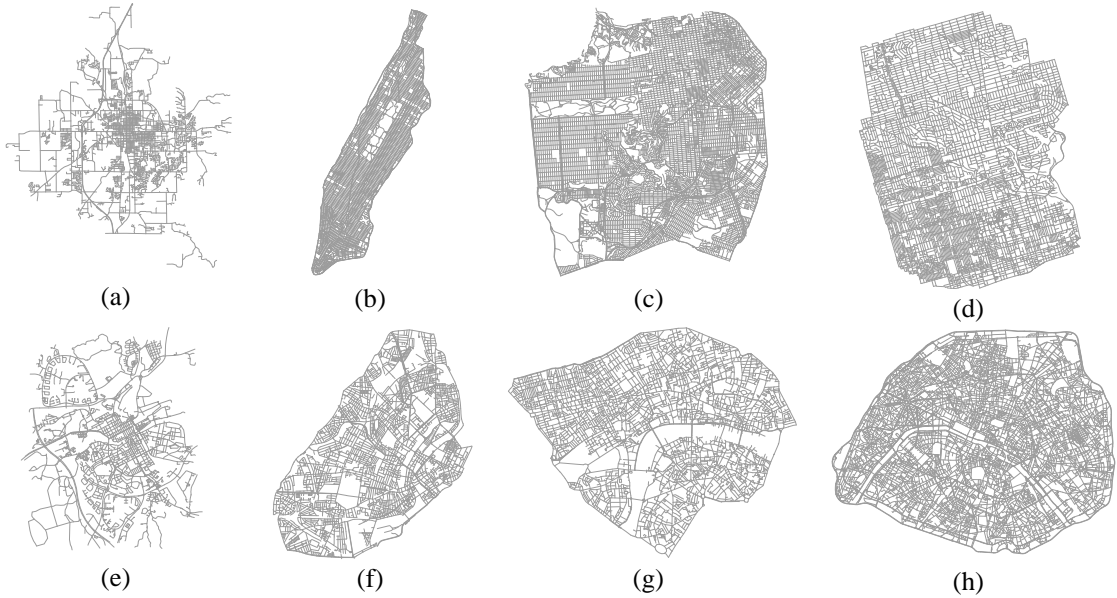

Figure 4: Maps of the six street networks, (a) Bloomington, (b) Manhattan, (c) San Francisco, (d) Toronto, (e) Gävle, (f) Copenhagen, (g) London, and (h) Paris

Table 2: Comparison between our solutions (FT and FTS) and existing solutions (ST, SP, and GMP) (Note: FT = fewest-turn paths, FTS = fewest-turn-and-shortest paths, ST = shortest paths, SP = simplest paths, GMP = Google Maps routes; x indicates comparison results in terms of both distance and turns; refer to Table 3 and 4 for the results)

|     | ST | SP | GMP |
| --- | --- | --- | --- |
| FT  | x  | x  | x   |
| FTS | x  | x  | x   |

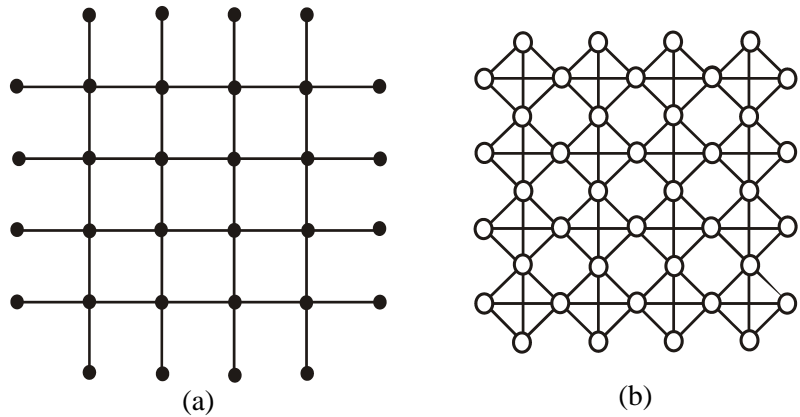

Figure 5: Illustration of geometry oriented graph (a) and its dual graph (b)

In the experiments, we adopt two parameters distance and the number of turns as a benchmark. For convenience, the distance (D) is denoted as a relative value (percentage), while the number of turns (T) are absolute values (see Table 3 and 4). We adopted the same exhausted approach as in Duckham and Kulik (2003) to derive routes from all points to all other points with respect to their dual graphs (Figure 5). In other words, both the simplest paths and the shortest paths were computed relying on the dual graphs rather than geometry oriented graphs. Because of this, the computation is very intensive. For the case of Bloomington, it implies that we had to compute the shortest, simplest, fewest-turn and fewest-turn-and-shortest paths for in total 4696 * 4696 pairs of points for the comparison purpose. The time cost ranges from a few hours for small networks like Bloomington, to dozens of hours for large networks like Paris. The computation was done in a desktop PC with the configuration of Intel Core QuadCPU, 3.25G RAM, and 1.0 TB Hard disk. It should be noted in the mean time that the implementation is very efficient for the computation can be instantly done without a noticeable delay for any pair of points and for any kind of routes. This all-to-all exhausted approach is not applicable to Google Maps, which restricts frequent request to its API. In this circumstance, we wrote a script to obtain the Google Maps routes with less frequent requests, and randomly chose 3000 routes from all-to-all routes for the comparison purpose. The results shown in Table 4 are based on the sample of 3000 points – a randomly chosen subset from all-to-all full set.

Now let us take a detailed look at the experimental results. Taking the second row in Table 3 as an example (of course using our comparison approach), we found that the fewest-turn paths are on average 6.3% shorter than the corresponding the simplest paths, and the number of turns is 1 less. This is a very encouraging result. Even more encouraging is the comparison between the fewest-turn-and-shortest paths and the simplest paths. The fewest-turn-and-shortest paths are 10.8% shorter than the corresponding simplest paths, while the number of turns is almost at the same level. In comparison with the shortest paths, the fewest-turn paths and the fewest-turn-and-shortest paths are respectively 18% and 12.7% longer, but the number of turns is dramatically reduced (only half). In the mean time, the simplest paths are 26% longer than the shortest paths, rather than 16% as previously reported by Duckham and Kulik (2003). Manhattan has even more encouraging results, but the results for the European cites are less encouraging. However, in all cases, the fewest-turn-and-shortest paths are always shorter than the simplest paths, while the number of turns is 1 less.

Table 3: Comparison results with SP and ST in terms of distances (D) and the number of turns (T) (Note: FT = fewest-turn paths, FTS = fewest-turn-and-shortest paths, ST = shortest paths, SP = simplest paths)

|  | FT/SP | | | FT/ST | | | FTS/SP | | | FTS/ST | | | SP/ST | | |
|---|---|---|---|---|---|---|---|---|---|---|---|---|---|---|---|
|  | D(%) | T(/) | | D(%) | T(/) | | D(%) | T(/) | | D(%) | T(/) | | D(%) | T(/) | |
| Bloomington* | 2.4 | 3.9 | 5.0 | 21.4 | 3.9 | 8.2 | -5.0 | 5.0 | 4.9 | 12.6 | 5.0 | 8.2 | 23.5 | 5.0 | 8.2 |
| Bloomington | -6.3 | 3.9 | 5.0 | 18.0 | 3.9 | 8.2 | -10.8 | 5.0 | 4.9 | 12.7 | 5.0 | 8.2 | 26.0 | 5.0 | 8.2 |
| Manhattan | -14.0 | 2.4 | 6.9 | 17.1 | 2.4 | 7.3 | -20.1 | 2.9 | 6.7 | 8.2 | 2.9 | 7.3 | 34.9 | 6.7 | 7.3 |
| San Francisco | 1.3 | 3.6 | 5.6 | 30.8 | 3.6 | 13.5 | -5.6 | 4.6 | 5.7 | 21.5 | 4.6 | 13.6 | 28.7 | 5.7 | 13.6 |
| Toronto | 7.8 | 3.6 | 5.1 | 21.9 | 3.6 | 12.9 | -1.3 | 4.0 | 5.2 | 10.6 | 4.0 | 12.9 | 12.1 | 5.2 | 12.9 |
| Gävle | 4.9 | 4.9 | 5.6 | 22.8 | 4.9 | 8.7 | -2.7 | 6.4 | 5.8 | 13.6 | 6.4 | 9.0 | 16.5 | 5.6 | 8.7 |
| Copenhagen | 5.1 | 3.6 | 5.1 | 31.9 | 3.6 | 9.0 | -4.1 | 4.8 | 5.2 | 20.0 | 4.8 | 9.1 | 25.2 | 5.2 | 9.1 |
| London | 17.1 | 4.7 | 6.4 | 48.3 | 4.7 | 10.6 | -7.0 | 5.9 | 6.5 | 17.0 | 5.9 | 10.7 | 25.8 | 6.5 | 10.7 |
| Paris | 2.6 | 4.6 | 7.4 | 53.1 | 4.6 | 12.0 | -16.9 | 5.8 | 7.4 | 23.3 | 5.8 | 12.1 | 48.4 | 7.4 | 12.1 |
| MEAN | 2.3 | 3.9 | 5.9 | 30.5 | 3.9 | 10.3 | -8.6 | 4.9 | 5.9 | 15.9 | 4.9 | 10.4 | 27.2 | 5.9 | 10.3 |

Table 4: Comparison results with GMP in terms of distances (D) and the number of turns (T)
(Note: FT = fewest-turn paths, FTS = fewest-turn-and-shortest paths, ST = shortest paths, GMP = Google Maps routes)

|  | FT/GMP | | | FTS/GMP | | | GMP/ST (FTS) | | | GMP/ST (FT) | | |
|---|---|---|---|---|---|---|---|---|---|---|---|---|
|  | D(%) | T(/) | | D(%) | T(/) | | D(%) | T(/) | | D(%) | T(/) | |
| Bloomington | -18.2 | 3.9 | 7.9 | -26.1 | 5.0 | 7.9 | 55.6 | 7.9 | 8.2 | 55.5 | 7.8 | 8.2 |
| Manhattan | -22.3 | 2.4 | 5.9 | -28.9 | 2.9 | 6.1 | 52.3 | 6.1 | 7.3 | 52.4 | 5.9 | 7.3 |
| San Francisco | -19.7 | 3.6 | 9.7 | -19.8 | 4.9 | 0.7 | 29.1 | 5.6 | 13.5 | 63.4 | 9.7 | 13.5 |
| Toronto | -15.8 | 3.6 | 6.6 | -24.0 | 4.0 | 6.6 | 45.7 | 6.6 | 12.9 | 46.3 | 13.3 | 12.9 |
| Gävle | 4.2 | 4.9 | 8.5 | -17.9 | 6.4 | 8.2 | 40.7 | 8.2 | 9.0 | 20.7 | 8.5 | 8.7 |
| Copenhagen | 16.9 | 3.6 | 11.8 | 6.2 | 4.8 | 11.8 | 12.4 | 11.8 | 9.1 | 13.2 | 11.8 | 9.0 |
| London | 20.4 | 4.7 | 11.4 | -5.5 | 5.9 | 11.2 | 23.7 | 11.2 | 10.7 | 24.5 | 11.4 | 10.6 |
| Paris | 16.1 | 4.6 | 17.6 | -6.5 | 5.8 | 17.7 | 31.1 | 17.7 | 12.1 | 32.2 | 17.6 | 12.0 |
| MEAN | -2.3 | 3.9 | 9.9 | -15.3 | 5.0 | 8.8 | 36.3 | 9.4 | 10.4 | 38.5 | 10.8 | 10.3 |

In comparison with the Google Maps routes, the fewest-turn paths are on average 2.3% shorter, while the number of turns is half as much. It is important to note that the -2.3% average makes little sense since percentages deviate substantially from case to case. For example, the Manhattan fewest-turn paths are 22.3% shorter than those of Google Maps, while for London, the fewest-turn paths are 20.4% longer than those of Google Maps. The fewest-turn-and-shortest paths are 15.3% shorter than those of Google Maps, while the number of turns is 3.8 less. In the mean time, we also made comparison between the Google Maps routes and the shortest paths, and found that the Google Maps routes are over 30% longer, and the number of turns is very similar. It is important to note that the comparison results with the Google Maps routes are not as accurate as the results with the simplest paths. The reason is that the comparison with the simplest paths is based on the same OSM data set, while comparison with the Google Maps routes is based on a different data set – the Google Maps routes on the Google Maps dataset and the fewest-turn and fewest-turn-and-shortest paths on OSM dataset. However, it is equally important to point out that results are reliable since motor ways are not much different between the Google Maps dataset and the OSM dataset.

In summary, the average distance of the fewest-turn paths is almost at the same level as the simplest paths, but the number of turns is much less. On the other hand, the average distance of the fewest-turn-and-shortest paths is much shorter than that of the simplest paths, while the number of turns is almost the same. It implies that both the fewest-turn and fewest-turn-and-shortest paths are superior to the simplest paths either for the distance or the turns. This superiority is even more obvious when compared to those of Google Maps for both the fewest-turn and fewest-turn-and-shortest paths. We believe that there are still possibilities to refine our algorithms, so that the distance of the fewest-turn routes can be further reduced. It is particularly true when our algorithms take into consideration of different morphological structures of road networks. All these are set our future work.

**5. Conclusion**
We developed a novel approach to computing fewest-turn routes or map directions based on the connectivity of natural roads. This approach is fundamentally different from the simplest paths algorithm, an algorithm essentially based on the connectivity of road segments for the computation. Our approach shows some striking advantages since the turn information is directly derived from the connectivity of natural roads. Not only topological information, but also semantic information in named streets is employed for computing the map directions. Thus our solution is more effective and comprehensive in terms of integration of all kinds of available information of road networks. Experiments demonstrated that the solution we provided is superior to the simplest paths and Google Maps routes from the perspective of distance and the number of turns involved. For example, the fewest-turn-and-shortest routes are on average 15% shorter than the routes suggested by Google Maps, while the number of turns is just half as much. Computationally, our approach is more efficient, for the connectivity graph is substantially smaller than the graph encoding the connectivity of road segments.

**Acknowledgement**
We thank Matt Duckham for kindly providing the dataset of Bloomington and discussions for the comparison study. We claim we had the equal contributions to the joint work that we are very proud of. Two anonymous referees deserve our special thanks for their constructive comments.

**Appendix: A short note on statistical comparison of two variables**

Given a series of pairs of the simplest paths $sp_i$ and the shortest paths $st_i$, how do we decide statistically which one is longer than another by a certain percentage? Duckham and Kulik (2003) adopted the following equation:

$$\alpha = \overline{\left(\frac{sp_i - st_i}{st_i}\right)} \tag{A1}$$

Where $\alpha$ is the ratio or percentage, and the bar denotes the average operator. If $\alpha$ is positive, then $sp$ is on average greater than $st$ by $\alpha$; otherwise $sp$ is less than $st$ by $\alpha$,

As we can see, Equation 1 first takes the ratio one by one, and then averages all the ratios. We think it is a problem to decide which one is longer. Instead, we suggest an equation that takes the averages first, and then the ratio as follows:

$$\beta = \frac{\overline{sp_i} - \overline{st_i}}{\overline{st_i}} \tag{A2}$$

Where $\beta$ is the ratio or percentage, and the bar denotes the average operator. In the same fashion, as in Equation A1, a positive $\beta$ indicates that $sp$ is greater than $st$ by $\beta$, and a negative $\beta$ indicates otherwise.

To illustrate the problem of Equation A1, let us generate 2 random variables (A1-A10, B1-B10) in Excel as shown in Table A1, and decide which one is bigger: A or B and by how much. According to Equation A1, we take the ratio first Ci = (Ai-Bi)/Bi, and then average AVERAGE(C1:C10) = 50% as shown in C11. The result indicates that A is bigger than B by 50%. On the other hand, according to Equation A2, we take averages first A12 = AVERAGE(A1:A10) and B12 = AVERAGE(B1:B10), and then ratio (A12-B12)/B12 = -7% as shown in C12. The result derived from Equation A2 simply says that A is less than B by 7%. Surprisingly, we end up with two contradictory results respectively from Equation A1 and A2.

Table A1: Comparison of two variables A and B in Excel – a simple example

|    | A        | B       | C     |
|----|----------|---------|-------|
| 1  | 132053   | 102110  | 0.29  |
| 2  | 311774   | 309624  | 0.01  |
| 3  | 273472   | 457325  | -0.40 |
| 4  | 89089    | 370157  | -0.76 |
| 5  | 376259   | 246054  | 0.53  |
| 6  | 102491   | 363935  | -0.72 |
| 7  | 415971   | 197516  | 1.11  |
| 8  | 142517   | 26237   | 4.43  |
| 9  | 308530   | 186460  | 0.65  |
| 10 | 312979   | 378142  | -0.17 |
| 11 |          |         | 0.50  |
| 12 | 246513.5 | 263756  | -0.07 |

It seems absurd to claim A is bigger than B by 50%. In fact, on average A is less than B: see A12 and B12. If one is still not convinced to this point, we can do a further calculation. Now that A is bigger than B by 50%, let's do the sum of (Bi + Bi * 50%) to see how it is equal (or close) to the sum of Ai. Unfortunately, the two sums are far different. On the other hand, AVERAGE(A1:A10) is far different from AVERAGE(B1:B10) + AVERAGE(B1:B10) * 50%. One can do the same comparison by replacing 50% by -7%, we will note that the sum of (Bi - Bi * 7%) is very close to the sum of Ai, or AVERAGE(A1:A10) is very close to AVERAGE(B1:B10) - AVERAGE(B1:B10) * 7%.